# Tilt-invariant scanned oblique plane illumination microscopy for large-scale volumetric imaging

MANISH KUMAR, AND YEVGENIA KOZOROVITSKIY*

*Department of Neurobiology, Northwestern University, Evanston, IL 60208, USA*
*Corresponding author: yevgenia.kozorovitskiy@northwestern.edu*

**This Letter presents the first demonstration of multi-tile stitching for large scale 3D imaging in single objective light-sheet microscopy. We show undistorted 3D imaging spanning complete zebrafish larvae, and over 1 mm$^3$ volumes for thick mouse brain sections. We use remote galvo scanning for light-sheet creation and develop a processing pipeline for 3D tiling across different axes. With the improved one photon (1p) tilt-invariant scanned oblique plane illumination (SOPi, /sōpī/) microscope presented here, we demonstrate cellular resolution imaging at depths exceeding 330 μm in optically scattering mouse brain samples, and dendritic imaging in more superficial layers.**

Light-sheet fluorescence microscopy (LSFM) has emerged as an indispensable tool in biology [1]. Among many developments in LSFM, the rise of single microscope objective based tilted light-sheet techniques for rapid volumetric imaging [2-4] holds promise for applications in neurobiology, where fast imaging of neuronal activity and other dynamics is critical. These new techniques overcome the limited steric access to the sample associated with classical LSFM arrangements [5]. Single objective based approaches use varied methods for remote scanning of the tilted light-sheet. OPM [3] relies on a piezo-electric actuator for on-axis movement of a remote objective, while SCAPE [4] uses a polygon scan mirror arrangement. Although the SCAPE approach simplifies the scanning architecture of OPM, it suffers from a scan position dependent tilt of the light-sheet.

To overcome the existing drawbacks of single objective based light-sheet microscopy techniques, we recently introduced SOPi [6], which uses a single planar scan mirror to provide tilt-invariant scanning of the oblique light-sheet. Figure 1(a) illustrates the comparison of OPM, SCAPE, and SOPi light-sheet scanning orientations. In previous work, we showed that this tilt-invariant scanning of SOPi is crucial for true perspective, 3D imaging of samples. However, like all other single objective light-sheet approaches, the SOPi implementation focused on imaging small sample volumes. No prior single objective light-sheet approaches have attempted to utilize the advantage of steric access to perform large-scale volumetric imaging of samples.

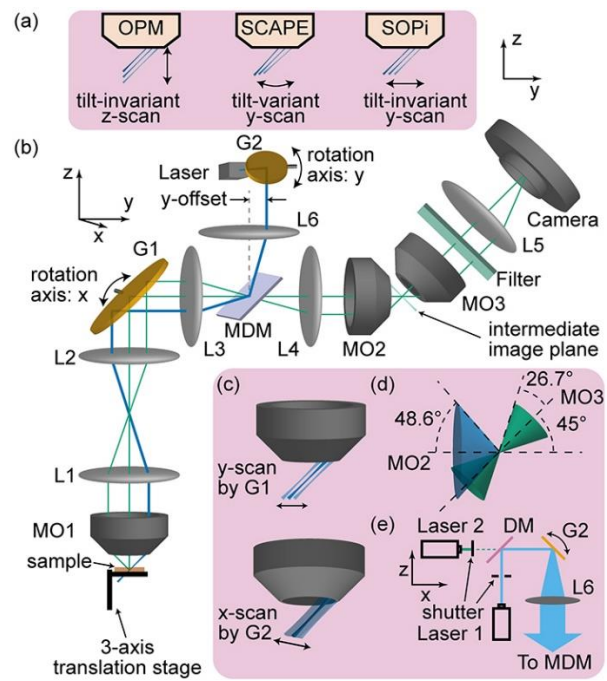

**Fig. 1.** (a) Comparison of OPM, SCAPE, and SOPi light-sheet scanning. (b) Schematics for the experimental setup of SOPi. (c) Role of G1 and G2. (d) Effective acceptance angle of the system, and (e) extended schematics showing laser arrangement for two color imaging. MO: microscope objective, G: galvo scanner, L: lens, MDM: multiband dichroic mirror.

In this Letter, we show for the first time that single objective based light-sheet microscopy can be used to image larger samples through multiple tile volume stitching. This is enabled by utilizing SOPi's unique tilt-invariant scanning geometry. Towards the goal of attaining large volume imaging capability, we also modify the 1p-SOPi system to image deeper in optically scattering media. This makes 1p-SOPi a preferred choice over 2p-SOPi in many applications where live samples could be damaged by the high illumination power needed for oblique light-sheet 2p imaging. High power requirements stem from relatively low 2p fluorescence cross-sections, as well as the low NA illumination necessary for light-sheet creation. Compared to our original 1p SOPi (SOPi 1.0) implementation [6], here we have 1) changed the light-sheet generation architecture to improve imaging performance, 2) incorporated two lasers to perform two color imaging, and 3) replaced the sCMOS camera with a low cost CMOS camera, expanding system utility. We also compare imaging quality of 1p SOPi 1.0 and SOPi 2.0, present a processing pipeline for 3D stitching, provide examples of stitched SOPi acquired volume-tiles, and describe how to obtain true perspective 3D visualization in stitched datasets.

The SOPi system working principles and optical layout have been described in detail elsewhere [6]. As has been done in conventional LSFM, we modify the 1p SOPi implementation by incorporating a DSLM scanning approach for light-sheet generation [7]. The DSLM approach reduces optical aberrations in the beam due to the lack of apertures, providing better optical sectioning capability. The system is arranged as shown in the schematics of Fig. 1(b). Two galvo scanners G1 (QS-12, 10 mm aperture, Nutfield Technology) and G2 (GVSM001, Thorlabs) are arranged so their rotation axes lie in conjugate planes of one another. Furthermore, the rotation axis of G1 is in conjugate plane to the back-focal planes (BFP) of both MO1 (20x, NA 1.0W, XLUMPLFLN20XW, Olympus) and MO2 (20x, NA 0.75, UPLSAPO20X, Olympus). This arrangement ensures that rotation of G1 and G2 provides tilt-invariant scanning as represented in Fig. 1(c). The illumination unit, unlike in the previous SOPi implementation, consists of laser 1 (473 nm, DPSS laser, Dragon Lasers) and laser 2 (532 nm, DJ532-40, Thorlabs), combined and co-aligned through a dichroic beam-splitter (FF495-Di03, Semrock). Fast scanning of G2 creates a light-sheet as shown in Fig. 1(e). A multiband dichroic mirror (Di03-R405/488/532/635, Semrock) reflects the illumination beam towards the sample and allows emitted fluorescence to pass. The amount of y-offset remains ~3.54 mm, corresponding to a 45° tilt [6].

The choice of converging lenses L1-L6 determines the effective magnification of the system. These must be chosen carefully, so that 1) the lateral and axial magnifications at the intermediate image plane in front of MO2 are equal to the ratio of the refractive indices of MO1 and MO2 immersion media; and 2) the overall system magnification provides Nyquist sampling. The first requirement minimizes optical aberrations while imaging an oblique plane [2,8], whereas the second constraint optimizes resolution and field of view of the system. We used achromatic doublet lenses from Thorlabs with focal lengths f = 200 mm (L1), f = 100 mm (L2, L3, and L6), f = 150 mm (L4), and f = 80 mm (L5). The value of the focal length of L5 was decided based on MO3 (20x, NA 0.45, LUCPLFLN20X, Olympus), the SOPi system's effective NA, and the camera pixel size (5.86 μm, GS3-U3-23S6M-C, FLIR). The system NA can be calculated from the effective overlap of acceptance cones [Fig. 1(d)]. The system NA = 1.33×sin((26.7°+48.6°-45°)÷2) ≈ 0.34. The system's effective magnification is ~11.8, providing Nyquist sampling of ~0.5 μm/pixel. This allows for a large field of view (here, ~950 μm along the x-axis). During imaging experiments, we used a 3-axes manual translation stage (PT3, Thorlabs) to position the sample within the field of view of SOPi system, and a custom MATLAB GUI to send ramp voltage signals to galvo scanners via a data acquisition card (PCIe-6321, National Instruments). μManager was used for camera control and image acquisition [9].

Here, a single sweep of oblique light-sheet acquires an image stack corresponding to a sheared cuboid shaped volume [Fig. 2(a)], with its edges predictably misaligned relative to the translation stage Cartesian coordinates (x,y,z). During processing of the image stack, Fiji/ImageJ [10,11] and other 3D reconstruction software manage the data in alternate coordinates (x',y',z'), so the default 3D volume representation [left, Fig 2(b)] is incorrect. Nevertheless, the reconstructed volume retains co-linearity due to the tilt-invariant scanning of SOPi. Therefore, the exact volume can be reconstructed by two simple geometrical transformations of scaling and shearing, as described in Fig. 2(b). In practice, we use a single 4×4 affine transformation matrix [6] to produce the combined geometrical transformation using the transformJ plugin [12].

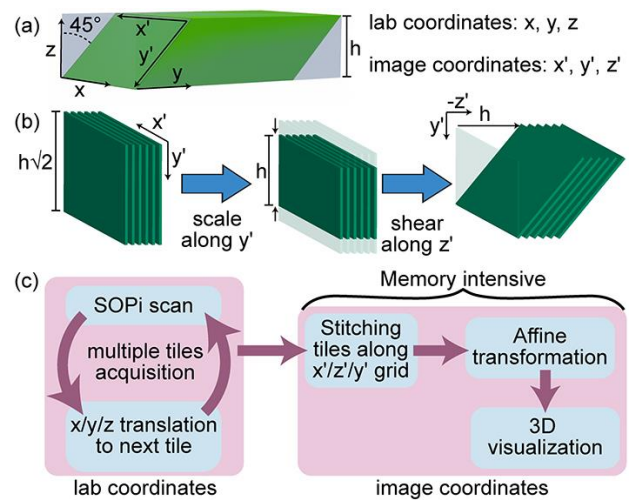

**Fig. 2**. (a) Relative orientation of sample (light grey cuboid), and SOPi acquired tile (green sheared cuboid). (b) Geometrical transformations to reshape tile into correct 3D orientation. (c) Processing pipeline for acquiring, stitching, and 3D visualization of multiple SOPi tiles.

Next, we investigate the imaging performance improvement due to the changes made in 1p SOPi. For this we used 1 mm-thick, 4% PFA (paraformaldehyde) fixed, uncleared coronal Thy1-GFP mouse brain sections through the hippocampus (007788, Jackson Laboratory). We imaged the same region of the sample, using two systems: first, the 1p SOPi 1.0 setup with a laser diode, slit-aperture, and cylindrical lens for light-sheet generation [Fig. 3(a)]; second, using the DSLM based [Fig. 1(e)] SOPi 2.0 presented here. We held the beam width (illumination NA) and power (0.55-0.6 mW) constant for both illumination approaches. A 3D perspective view of the scanned ~400 μm × 400 μm × 160 μm (x'× y'× y) volume is presented in Fig. 3(b), 3(c) and Visualization 1. Many processes are resolved in Fig. 3(c) compared to Fig. 3(b), and feature continuity is improved. For a quantitative comparison on features present in both data sets, we plotted intensity profiles [Fig. 3(f)] of a line segment along the y'-axis in one x'y' section [Fig. 3(d), 3(e)] of the sample. The cell body marked by arrows in Fig. 3(d), 3(e) was used for intensity normalization. Smaller scattering background for a given signal across depth (y′-axis) with SOPi 2.0 is apparent in Fig. 3(f) (average % increase in SBR at superficial depth and deeper regions, 32 and 10, for n=70 measurements).

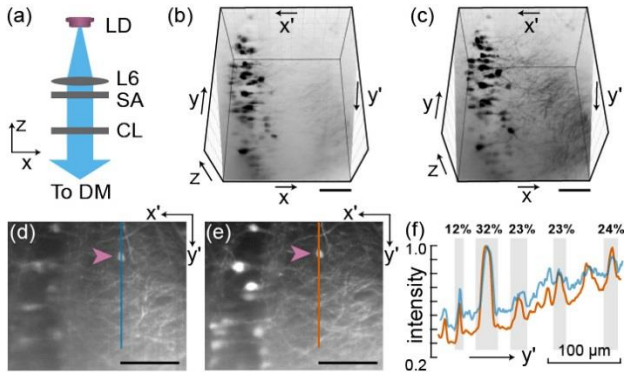

**Fig. 3**. Comparison of SOPi 1.0 and 2.0. (a) Light-sheet generation in the original SOPi 1.0. A 3D perspective view (inverted grey LUT) of the same scanned volume using SOPi 1.0 in (b), and SOPi 2.0 in (c). Also see Visualization 1. View of an oblique (x′y′) section of the sample using SOPi 1.0 in (d) and SOPi 2.0 in (e). The arrows point to the cell body used for normalizing the intensity plot shown in (f). Vertical line segments, corresponding to the intensity plots, are marked in (d) and (e). Features shaded in grey illustrate higher signal to background ratio across depth (% increase as noted). LD: laser diode, SA: slit aperture, CL: cylindrical lens, DM: dichroic mirror. (Scale bar: 100 μm)

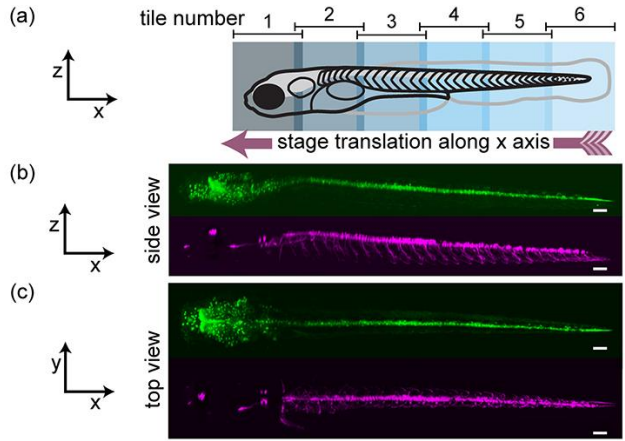

**Fig. 4**. Stitching multiple tiles along the x-axis. (a) Schematics of tile arrangement along the length of the fish. 3D perspective side view in (b) and top view in (c) of green and red fluorescence. Also see Visualization 2. (Scale bar: 100 μm)

With the 1p SOPi 2.0, we investigate how tilt-invariant scanning can be used for stitching tiles. Figure 2(b) illustrates that although the affine transformation places tiles in the correct orientation, tiles acquire corner padding (of blank pixels) rendering them un-stackable along the z′ direction. The simplest solution to this problem is to stitch raw tiles in their original form, *i.e.* pre-affine transformation, as depicted in Fig. 2(c) workflow. All tiles are acquired by moving translation-stage/sample and stitched together with either pairwise-stitching or BigStitcher to form large volume data [13,14]. Tiles acquired along x, y and z axes in lab coordinates are stitched along x′, z′ and y′ axes in image coordinates. A single operation of affine transformation on stitched volume data reshapes it into an exact 3D representation of sample volume. This large volume data is then visualized using BigDataViewer [15] or ClearVolume [16]. Note that no deconvolution or other post-processing is required, but could be implemented for further improvements in image quality.

Now, we present examples of stitching SOPi tiles along the x, y and z axes. In the first example, we stitch multiple tiles along the x-axis. For this, we used an agar gel embedded 4 dpf (days post fertilization) zebrafish larva from an *olig2:GFP* cross to *mnx:Gal4;UAS:pTagRFP*. The fish was oriented with its length along the x-axis, and a total of six overlapping tiles, each spanning ~0.9 mm × 0.6 mm × 0.4 mm (x′× y′× y), were acquired with manual translation of the stage to cover the zebrafish (4 mm long) as illustrated in Fig. 4(a). Each SOPi tile was acquired at 50 fps in 6 seconds with G2 driven at 100 Hz. In the processing pipeline, each tile was first scaled along the x′ and y′ direction to half (to reduce data size). Tiles were stitched with pairwise-stitching, affine transformed with transformJ, and visualized with ClearVolume plugins. 3D reconstruction of the entire zebrafish, for both red and green fluorescence, is presented in Fig. 4 and Visualization 2.

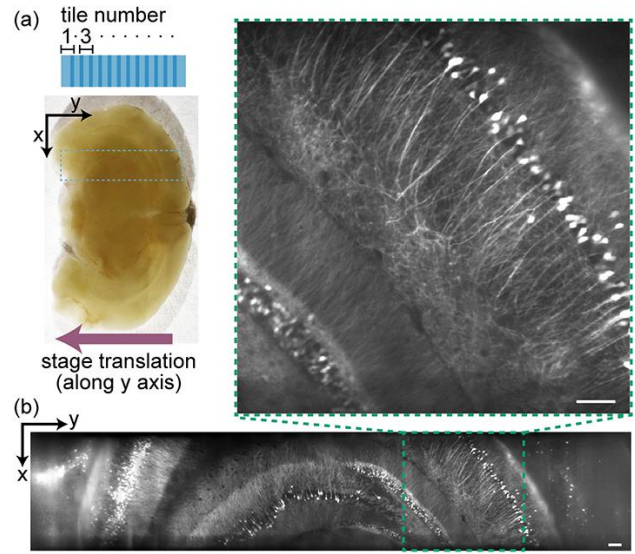

**Fig. 5**. Stitching multiple tiles along the y-axis. (a) Tile arrangement (top view) along a 1 mm thick, uncleared mouse brain section. Scanned region highlighted by the dashed rectangle. (b) A virtual slice from the stitched dataset, at the depth of 100 μm, along with the inset showing an enlarged view. Also see Visualization 3. (Scale bar: 100 μm)

In the second example, we show stitching of multiple tiles acquired along the y-axis. The brain section [Fig. 5(a)] was translated along the y-axis in steps of ~200 μm to cover ~4.75 mm length through multiple overlapping tiles. Each tile spanning ~950 μm × 435 μm × 250 μm (x′× y′× y) was acquired in 5 seconds at 50 fps. We restricted scan range to 250 μm for uniform illumination throughout the y-sweep. We used BigStitcher to stitch the tiles, transformJ to affine transform, and BigDataViewer to visualize the volume. The stitched volume exceeded 1 mm³, with dimensions of ~ 0.95 mm × 4.75 mm × 0.3 mm (x × y × z). Figure 5(b) shows a virtual xy slice from a stitched volume at the depth of 100 μm where all the cell bodies and many dendrites are clearly visible. Visualization 3 shows an oblique plane (x′y′) scan through the entire stitched volume.

In the final example, we demonstrate stitching along the z-axis. Figure 6(a) illustrates how two connected SOPi tiles along sample depth are acquired by moving the sample at 45° to both y and z axes. We acquired two overlapping tiles in the same mouse brain section by translating the sample diagonally by ~250 μm. Each tile was acquired at 50 fps, spanning 400 μm in 6 seconds. Tiles were stitched pairwise, affine transformed, and visualized with BigDataViewer. In this dataset, spanning ~950 μm × 400 μm × 400 μm (x × y × z), the depth penetration of SOPi becomes apparent. Neuronal cell bodies are visible at greater than 330 μm depth, with dendritic processes well-resolved at more superficial depths in an optically scattering mouse brain section [Fig. 6(b) and Visualization 4]. This depth performance exceeds any previously published single objective 1p light-sheet microscopy approach.

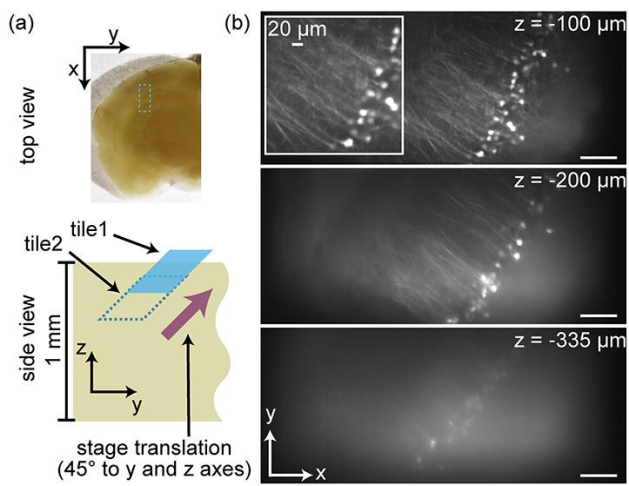

**Fig. 6**. Stitching tiles along sample depth (z-axis). (a) Placement and orientation of connected tiles along depth. The top view of the tile is highlighted by a rectangle. (b) Virtual xy slices along the depth of the stitched volume of a thick mouse brain section. Some neurons at >350 μm depth (see Visualization 4) are resolved, with neuronal processes imaged at more superficial depths. (Scale bar: 100 μm)

In conclusion, we have modified the 1p SOPi illumination architecture to image deeper in scattering samples. In addition, we obtain large-scale volumetric imaging by stitching multiple volume scans together. These advances make SOPi suitable for *in vivo* imaging in mice as well as other large biological samples. Moreover, the current implementation supports relatively high speed acquisition of high quality data with basic, inexpensive cameras. The use of high sensitivity sCMOS cameras would further speed up volume acquisitions [6]. Since scanning during a tile acquisition is done remotely with galvo scanners, there are no vibration artifacts induced during imaging. Thus, a manual translation stage, unlike in conventional light-sheet approaches, is sufficient for large volume stitching. With the help of an automated translation stage and a workstation for data processing, experiments can be significantly scaled up, *e.g.* to image multiple zebrafish or other large samples in parallel. In the future, the use of self-reconstructing beams [17,18] with SOPi should provide much deeper imaging capabilities. The already available choices of higher NA objectives and post-processing algorithms would enable future SOPi implementations to image at sub-dendritic and potentially molecular resolution.

**Funding.** NIMH R01MH117111; Beckman Young Investigator Award; Searle Scholar Award; William and Bernice E. Bumpus Young Innovator Award; Rita Allen Foundation Scholar Award; Sloan Research Fellowship.

**Acknowledgment**. We thank Drs. S. Kishore and D. McLean for providing zebrafish samples; D. Badong and Dr. M. Priest for preparing mouse brain sections; E. Szuter for maintaining zebrafish colony; and L. Butler for mouse colony management.

**Disclosure.** A patent has been filed based on this work.

**References:**

1. R. M. Power and J. Huisken, "A guide to light-sheet fluorescence microscopy for multiscale imaging," Nat. Methods **14**(4), 360-373 (2017).
2. C. Dunsby, "Optically sectioned imaging by oblique plane microscopy," Opt. Express **16**(25), 20306–20316 (2008).
3. M. B. Sikkel, S. Kumar, V. Maioli, C. Rowlands, F. Gordon, S. E. Harding, A. R. Lyon, K. T. MacLeod, and C. Dunsby, "High speed sCMOS-based oblique plane microscopy applied to the study of calcium dynamics in cardiac myocytes," J. Biophotonics **9**(3), 311–323 (2016).
4. M. B. Bouchard, V. Voleti, C. S. Mendes, C. Lacefield, W. B. Grueber, R. S. Mann, R. M. Bruno, and E. M. Hillman, "Swept confocally-aligned planar excitation (SCAPE) microscopy for high speed volumetric imaging of behaving organisms," Nat. Photonics **9**(2), 113–119 (2015).
5. P. G. Pitrone, J. Schindelin, L. Stuyvenberg, S. Preibisch, M. Weber, K. W. Eliceiri, J. Huisken, and P. Tomancak, "OpenSPIM: an open-access light-sheet microscopy platform," Nat. Methods **10**(7), 598–599 (2013).
6. M. Kumar, S. Kishore, J. Nasenbeny, D.L. McLean, and Y. Kozorovitskiy, "Integrated one- and two-photon scanned oblique plane illumination (SOPi) microscopy for rapid volumetric imaging," Opt. Express **26**(10), 13027-13041 (2018).
7. P. J. Keller, A. D. Schmidt, J. Wittbrodt, and E. H. K. Stelzer, "Reconstruction of zebrafish early embryonic development by scanned light sheet microscopy," Science **322**(5904), 1065–1069 (2008).
8. E. J. Botcherby, R. Juskaitis, M. J. Booth, and T. Wilson, "Aberration-free optical refocusing in high numerical aperture microscopy," Opt. Lett. **32**(14), 2007–2009 (2007).
9. A. D. Edelstein, M. A. Tsuchida, N. Amodaj, H. Pinkard, R. D. Vale, and N. Stuurman, "Advanced methods of microscope control using μManager software," J. Biol. Methods **1**(2), e10 (2014).
10. J. Schindelin, I. Arganda-Carreras, E. Frise, V. Kaynig, M. Longair, T. Pietzsch, S. Preibisch, C. Rueden, S. Saalfeld, B. Schmid, J. Y. Tinevez, D. J. White, V. Hartenstein, K. Eliceiri, P. Tomancak, and A. Cardona, "Fiji: an open-source platform for biological-image analysis," Nat. Methods **9**(7), 676–682 (2012).

11. C. A. Schneider, W. S. Rasband, and K. W. Eliceiri, "NIH Image to ImageJ: 25 years of image analysis," Nat. Methods **9**(7), 671–675 (2012).
12. E. H. W. Meijering, W. J. Niessen, and M. A. Viergever, "Quantitative Evaluation of Convolution-Based Methods for Medical Image Interpolation," Med. Image Anal. **5**(2), 111–126 (2001).
13. S. Preibisch, S. Saalfeld, and P. Tomancak, "Globally optimal stitching of tiled 3D microscopic image acquisitions," Bioinformatics, **25**(11), 1463-1465 (2009).
14. D. Hörl, F.R. Rusak, F. Preusser, P. Tillberg, N. Randel, R. K. Chhetri, A. Cardona, P. J. Keller, H. Harz, H. Leonhardt, M. Treier and S. Preibisch, "BigStitcher: Reconstructing high-resolution image datasets of cleared and expanded samples," bioRxiv, p.343954 (2018).
15. T. Pietzsch, S. Saalfeld, S. Preibisch, and P. Tomancak, "BigDataViewer: visualization and processing for large image data sets," Nat. Methods, **12**(6), 481-483 (2015).
16. L. A. Royer, M. Weigert, U. Günther, N. Maghelli, F. Jug, I. F. Sbalzarini, and E. W. Myers, "ClearVolume: open-source live 3D visualization for light-sheet microscopy," Nat. Methods **12**(6), 480–481 (2015).
17. F. O. Fahrbach, P. Simon, and A. Rohrbach, "Microscopy with self-reconstructing beams," Nat. Photonics **4**(11), 780-785 (2010).
18. T. Vettenburg, H. I. Dalgarno, J. Nylk, C. Coll-Lladó, D. E. Ferrier, T. Čižmár, F. J. Gunn-Moore, and K. Dholakia, "Light-sheet microscopy using an Airy beam," Nat. Methods **11**(5), 541-544 (2014).